\documentclass[aps,pre,showpacs,twocolumn,superscriptaddress]{revtex4}
\usepackage{epsfig}
\usepackage{amsmath,amssymb,wasysym,epsfig,capt-of,ifthen,calc}
\usepackage{bbm}
\usepackage{latexsym}   %extra symbols
\usepackage{setspace}   %doublespacing
\usepackage{array}      %for some equations
\usepackage{delarray}   %for some equations
\usepackage{afterpage}
\usepackage{graphicx}% Include figure files
\usepackage{dcolumn}% Align table columns on decimal point
\usepackage{bm}% bold math
\usepackage{array}
\usepackage{hyperref}
\usepackage{float}
\usepackage{supertabular}
\usepackage{longtable}
\newcommand{\be}{\begin{equation}}
\newcommand{\ee}{\end{equation}}
\newcommand{\bea}{\begin{eqnarray}}
\newcommand{\eea}{\end{eqnarray}}

\usepackage{amsmath}

\bibstyle{apsrev}
\begin{document}

\title{Localization Transition of Biased Random Walks on Random Networks}

\author{Vishal Sood} 
\affiliation{Complexity Science Group, University of Calgary, Calgary, Canada}
\affiliation{Institute for Biocomplexity and Informatics, University of Calgary,
 Calgary, Canada}
\author{Peter Grassberger}
\affiliation{Complexity Science Group, University of Calgary, Calgary, Canada}
\affiliation{Institute for Biocomplexity and Informatics, University of Calgary,
 Calgary, Canada}

\date{\today}

\begin{abstract}
We study random walks on large random graphs that are biased towards a randomly 
chosen but fixed target node. We show that a critical bias strength $b_c$ exists
such that most walks find the target within a finite time when $b>b_c$. For $b<b_c$, 
a finite fraction of walks drifts off to infinity before hitting the target. The 
phase transition at $b=b_c$ is a critical point in the sense that quantities like 
the return probability $P(t)$ show power laws, but finite size behavior is complex 
and does not obey the usual finite size scaling ansatz. By extending rigorous 
results for biased walks on Galton-Watson trees, we give the exact analytical 
value for $b_c$ and verify it by large scale simulations.
\end{abstract}


\maketitle

Random walks are a fascinating subject, both for their intrinsic mathematical 
beauty and for their wide range of applications~\cite{hughes::RWRE}. One of
the most celebrated results is that unbiased random walks on regular lattices
are recurrent for dimension $d\leq 2$, while they are transient for 
$d>2$~\cite{polya::recurrence}. This means that a walk starting at a node 
${\bf x}$ certainly returns to ${\bf x}$ when $d\leq 2$, but has a 
finite chance to escape to infinity  when $d>2$.

Here we study an analogous problem for walks on random
graphs which lack small loops in the limit of infinite graph size. Examples
of such graphs include the Erd\"os-Renyi (ER) random graphs, as well as random
graphs with any fixed degree sequence, provided that the variance of the 
degree distribution is finite. We call the latter Molloy-Reed (MR) 
graphs~\cite{molloy::1995}. An ER graph with $N$ nodes is constructed 
by introducing a link between each pair of nodes with probability $p$. 
It is in many ways similar to an infinite dimensional lattice. In particular, 
its diameter $l$ increases only logarithmically with the number of nodes, 
while $N \sim l^d$ on a $d$-dimensional lattice. Thus one should expect 
unbiased random walks on an ER graph to be transient. Although less is known
rigorously about MR graphs, we expect the same to be true for them. In order 
to arrive at a non-trivial problem, we thus consider walks biased towards 
a randomly chosen but fixed ``target" node. We show that there is a phase 
transition from recurrence (or {\it localization}) to de-localization at a 
critical bias strength. Notice that this is unrelated to similar phase 
transitions observed e.g. in \cite{dhar::1984}, where the bias is not towards but
{\it away from} the target.

This seemingly abstract mathematical problem has a number of practical applications.
Consider, e.g., routing a message from node $A$ to node $B$ on the
internet. Each node $i$ maintains a routing table which 
indicates for each target node the optimal first step, parting 
from $i$. Using her routing table, $A$ sends the message to her optimal neighbor 
$i_1$. From there it is sent, using the routing table at $i_1$, to $i_2$, etc., 
until $i_n=B$ is reached. If all routing tables are correct and up-to-date, the 
message reaches its destination along the optimal path ({\it i.e.}, the path 
between $A$ and $B$ with the shortest length). However, some routing tables 
may be faulty, either because they contain mistakes or have become obsolete
due to changes in the internet topology. If the fraction of such nodes is below 
a certain threshold, the effect is small and routing is still efficient: the 
average time to go from $A$ to $B$ scales linearly with the distance. 
But when the fraction exceeds a critical value, the message might take a very long 
and convoluted path through a finite fraction of the entire network, before reaching 
its destination. There are of course many details in which routing on the real 
internet differs from the simpler problem of biased random walks on random 
graphs~\cite{willinger::criticality}. For instance, the degree distribution of the 
internet is approximately scale-free with a divergent second moment~\cite{albert::1999}; 
stray messages on the internet are killed after some time; and the bias is implemented 
differently and is quenched (the same routing tables are used at successive time 
steps). Despite these technical details, the two problems are basically the same. 
For a related discussion of communication based on noisy routing, see \cite{rosvall}.

Another application is to quantum mechanics. A random walker on a graph corresponds
quantum mechanically to a particle with hopping dynamics. The recurrence problem for 
an unbiased walk corresponds then to the question whether
such a particle, subjected to an attractive $\delta$-potential concentrated on 
a single node, forms a bound state. 
Localization of biased random walks corresponds to the existence of bound states 
for potentials which increase linearly with distance from the target node. 
Delocalization would imply the paradoxical
situation that no bound state exists, although the potential increases forever
as one goes further away from the target node. Instead, a particle released near
the bottom of the potential continues to climb up the potential forever, because
there are always more paths leading uphill than leading back to the bottom.

Technically, we define our walks as follows. Consider a finite but large undirected 
random graph ${\cal G}$ with $N$ nodes and with degrees chosen from a distribution 
$P_{\cal G}(k)$. We assume that $P_{\cal G}(k)$ is such that ${\cal G}$ is sparse
and has a giant connected component (GC). We define the distance between any two nodes 
as the number of links in the shortest path connecting them.  We randomly choose 
a node $A$ on ${\cal G}$ and label all other nodes according to their distance 
from $A$.  Consider node $i$ at distance $d_i$ from $A$, which has $k^-_i$, 
$k^0_i$ and $k^+_i$ neighbors at distances $d_i+1$, $d_i-1$, and $d_i$ from $A$, 
respectively.  The random walk steps from $i$ with probabilities
\be
   p_i^-=\frac{b}{{\cal N}_i}, \quad p_i^0=\frac{1}{{\cal N}_i},
 \quad p_i^+=\frac{b^{-1}}{{\cal N}_i}\;.
                   \label{jumps-ER}
\ee
to a node closer to, at the same distance as or further away from $A$, respectively.
Finally, normalization requires ${\cal N}_i = b k^-_i + k^0_i + b^{-1}k^+_i\;.$

Similar walks on regular lattices have been studied repeatedly, see e.g.
\cite{mehra::bias}. More importantly, Eq.~(\ref{jumps-ER}) is a generalization 
of the bias used in the `$\lambda$-biased random walks' studied on Galton-Watson 
(GW) trees \cite{peres::2006,lyons::96biased,lyons::90walks}. Starting at the root, 
each node on the tree has a number of daughter nodes chosen from a prescribed
distribution. The root is chosen as the target $A$ of the random walk. 
On a tree, every node $i\neq A$ has only one neighbor closer to the target, 
$k^-_i=1$, and no neighbors at the same distance, $k^0_i=0$. The probabilities 
of the next step are chosen such that~\cite{peres::2006,lyons::96biased}
$ p_i^- / p_i^+ = \lambda$. This corresponds to Eq.~(\ref{jumps-ER}), restricted 
to tree-like graphs, with $b = \sqrt{\lambda}$.

The typical length of a loop in an undirected graph ${\cal G}$ with finite 
mean degree and finite degree variance is of order ${\ln N}$
{\cite{bollobas::MGT,janson::RG}}. If only local properties are of interest, 
the graph can, as $N \rightarrow \infty$, be effectively
replaced by a GW tree.  However, 
a subtlety should not be overlooked: to obtain a rooted tree
${\cal T}$ from a loopless undirected graph ${\cal G}$, we have to 
choose randomly a node $A$ as the root of the tree, and draw an arrow on 
each link pointing away from $A$. A node with degree $k$ on ${\cal G}$ has
in-degree 1 and out-degree $k - 1$ on the effective tree. 
The out-degree distribution on the tree can be related to the
degree distribution $P_{\cal G} (k)$ of the graph {\cite{molloy::1995}},
%If the degree distribution on ${\cal G}$
%is $P_{\cal G} (k)$, then the average out-degree distribution on the
%tree, averaged over all nodes $A$, is {\cite{molloy::1995}}
\be
   P_{\cal T}^{\rm out} (k^+) = \frac{k}{\langle k\rangle}
   P_{\cal G} (k), \label{treeDegreegraph}
\ee
with $k=k^++1$. The prefactor $k / \langle k \rangle$ on the right hand side 
takes into account the fact that each of the $k$ links attached to the node 
can play, with the same probability, the role of the incoming link on the 
tree. 

A GW tree will grow to infinity only if its average
out-degree is larger than $1$~\cite{athreya::BP}.
This implies that the underlying random graph is connected when
{\cite{molloy::1995,cohen::2000}}
\be
  \mu = \langle k^+ \rangle_{{\rm GW}} = 
    \langle k (k - 1) \rangle_{\mathcal{G}} / \langle k
    \rangle_{\mathcal{G}} > 1, \label{RGinftyBR} 
\ee
which we  assume to be satisfied. Further for the graph to be sparse
and lack small loops we assume 
that $\mu \ll N$.

If there is a critical bias $\lambda_c$ for walks on a GW tree so that walks
are localized near the root for $\lambda > \lambda_c$ and delocalized for
$\lambda < \lambda_c$, then an analogous critical bias $b_c \leq 
\sqrt{\lambda_c}$ must also exist for graphs. The reason is simply that
only local properties are relevant when the walks are localized
\footnote{Notice that this argument is not strictly correct, since it 
neglects the exponentially small probability for delocalized walks even 
when $\lambda > \lambda_c$.}. Hence
localization on a GW tree implies localization on the corresponding graph.
Conversely, assume that $b_c$ were strictly smaller than $\sqrt{\lambda_c}$.
Then walks on the GW tree with $b_c^2 < \lambda < \lambda_c$ would also be
localized, contradicting the starting assumption that the critical point is
at $\lambda_c$. Thus we must have $b_c = \sqrt{\lambda_c}$.

Indeed, it is known that the $\lambda$--biased random walk on a GW tree has a
localization transition at $ \lambda_c = \mu$,
such that the walk is recurrent for $\lambda > \lambda_c$ and transient
for $\lambda < \lambda_c$ {\cite{lyons::90walks}}. For $\lambda = \lambda_c$ 
the authors of \cite{peres::2006} prove a central limit theorem which states
that the walk behaves, as far as the distance from the root is concerned, 
like unbiased 1-d Brownian motion. In the following we will summarize the 
arguments leading to those conclusions and indicate modifications required 
to apply to ER and MR graphs.

For a walker at a node $i$ that is different from the root $A$, we define $p_i^{\rm E}$
to be the probability that it escapes to infinity before it hits the root.
We denote by ${\cal P}(i)$ the set of parents of $i$, i.e. the neighbors
that are closer to $A$ than $i$. Similarly, ${\cal S}(i)$ is the set of 
siblings of $i$ (i.e. $d_j = d_i$ for all $j\in {\cal S}(i)$), and ${\cal C}(i)$
denotes its children. Finally, we define $p_A^{\rm E}=0$. Then we have, for 
any $i\neq A$,
\be
   p_i^{\rm E}= {1\over {\cal N}_i}\left[b\sum_{j \in {\cal P}(i)} p_j^{\rm E}
                                         +  \sum_{j \in {\cal S}(i)} p_j^{\rm E}
                                 +{1\over b}\sum_{j \in {\cal C}(i)} p_j^{\rm E}
                                 \right]\;.
\ee
Using ${\cal N}_i$ defined after Eq.~(\ref{jumps-ER}) we can rearrange this to
\be
  \sum_{j \in \mathcal{P} (i)} (p_i^{{\rm E}} - p_j^{{\rm E}}) =
  \frac{1}{b^2} \sum_{j \in \mathcal{C} (i)} (p_j^{{\rm E}} -
  p_i^{{\rm E}}) + \frac{1}{b^{}} \sum_{j \in \mathcal{S} (i)}
  (p_j^{{\rm E}} - p_i^{{\rm E}}).                \label{escdiff1} 
\ee
A sum over all $i$ with fixed $d_i=d$, cancels all the contributions of the siblings. 
Defining 
%$   X_d = \sum_{ i:\;d_i=d \atop j \in {\cal P}(i)} (p_i^{\rm E} - p_j^{\rm E})$
 \be 
  X_d = \sum_{ i:\;d_i=d \atop j \in {\cal P}(i)} (p_i^{\rm E} - p_j^{\rm E})\;,
 \ee
we get the recursion
\be
   X_d = \frac{1}{b^2} X_{d+1}\;.                    \label{escsum}
\ee                
which can be iterated to give the average escape probability from all
children of the root
\be
   p_{\rm E} \equiv \frac{1}{k_A} \sum_{i\in {\cal C}(A)} p_i^{\rm E} 
   = \frac{1}{k_Ab^{2d}} X_{d+1}\;.
                                                      \label{pescl} 
\ee
For trees this simplifies because each node (except for the root) has only
one parent, and the number of terms on the r.h.s. increases in average as 
$\mu^d$ for large $d$. Since each term is bounded, the total sum increases
at most as $(\mu/b^2)^d$. Thus $p_{\rm E}=0$ for $b^2>\mu$, showing that
the walk is recurrent. The proof that $b_c$ is not only $\leq \sqrt{\mu}$, 
but %that actually 
\be
   b_c = \sqrt{\mu} \;,     \label{bc}
\ee
is found in \cite{lyons::90walks}. Basically, one replaces in Eq.~(\ref{pescl})
the number of terms by its expected value and the difference 
$p_i^{\rm E} - p_j^{\rm E}$ by its average $\langle p \rangle_{d+1}
-\langle p \rangle_d$, to obtain
\be
   p_{\rm E} \sim \left( \frac{\mu}{b^2} \right)^d \left( \langle p
  \rangle_{d + 1} - \langle p \rangle_d \right)     \label{esceqn}
\ee
which has the solution
\be
   \langle p \rangle_d = \alpha - \beta \left( \frac{b^2}{\mu} \right)^d
\ee
with $\alpha$ and $\beta$ being constants. For $b^2>\mu$, the only 
consistent values are $\alpha=\beta=0$. For $b^2<\mu$ one has a non-zero
solution, indicating $p_{\rm E}>0$.

%The factor $(\mu/b^2)^d$ in Eq.~(\ref{esceqn}) compares the rate at which
%the graph grows with distance $d$ from the root and the effective bias at that distance.
For general finite MR graphs the number of nodes at distance $d$ from the root increases 
slower than $\mu^d$, due to the existence of loops. In this case 
$b_c\leq \sqrt{\mu}$ holds {\it a fortiori}, but the rigorous proof that also 
$b_c\geq \sqrt{\mu}$ becomes more difficult.
The corrections to $\mu^d$ are small for small $d$ (there are few small
loops), and we can expect that the effect of loops on $b_c$ can be 
neglected in the limit $N\to\infty$. But loops should effect the finite
size scaling behavior.

To verify the existence of a localization transition on large ER graphs, 
check Eq.~(\ref{bc}), and study finite size effects, we perform large scale 
numerical simulations. We first generate large ER graphs (with up to $2\times 10^8$ 
nodes), for several nominal values of $\langle k\rangle$. These are chosen such that 
the GC contains more than 90\% of all nodes. Pruning all nodes and links
not connected to the GC gives the final graph size $N$. This
procedure increases $\langle k\rangle$ slightly and makes 
the degree distribution slightly non-Poissonian, so that $\mu$ no longer coincides with 
the naive theoretical estimate $\mu = \langle k\rangle$ for ER graphs. Instead, it is 
estimated from the exact definition Eq.~(\ref{RGinftyBR}).

After extracting the GC, we choose randomly one of its nodes as the target 
$A$ and compute every other node's distance from $A$. Then we start $M$ walks
at $A$ and follow them until they return to $A$. This is repeated $M'$ times by 
taking new target nodes, and finally the whole procedure is repeated for $M''$ 
different graphs.

The first observable to be discussed is the average return time. 
For $b=1$, i.e. for unbiased random walks, the average return time $\langle t 
\rangle$ on a finite graph with $N$ nodes is $\propto N$ \cite{bollobas::MGT}, 
We 
expect the same to hold for all $1\leq b<b_c$. For the critical bias $b=b_c$,
since the distance of the biased walk from the root behaves as a
1-d Brownian motion, the return time should grow as ${\rm log}N$ which is
the diameter of the graph. In contrast, $\lim_{N\to\infty} \langle 
t\rangle <\infty$ in the localized regime $b>b_c$. 

\begin{figure}
  \begin{center}
  \epsfig{file=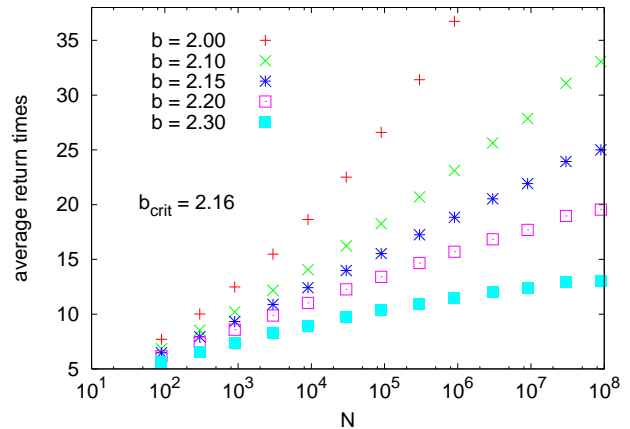, width=8.5cm, angle=0}
 \caption{(color online) Average return times against the system size $N$ for different
values of the bias $b$, on ER graphs with $\langle k\rangle = 14/3$. }
  \label{average_times}
  \end{center}
\end{figure}

In Fig.~\ref{average_times} we show $\langle t \rangle$ versus $N$ for various 
biases on a linear-log scale. The nominal number $L/N$ of links per node in 
constructing the ER graphs was $7/3$, which
would give $\mu = 14/3 = 4.6667$. For the GC, we found numerically 
$\mu = 4.6673$. The critical bias should be $b_c = 2.1604$,
which certainly agrees with 
with Fig.~\ref{average_times}.
% although the estimate of $b_c$ that we get 
%from Fig.~\ref{average_times} is not very precise. 
  
\begin{figure}
  \begin{center}
  \epsfig{file=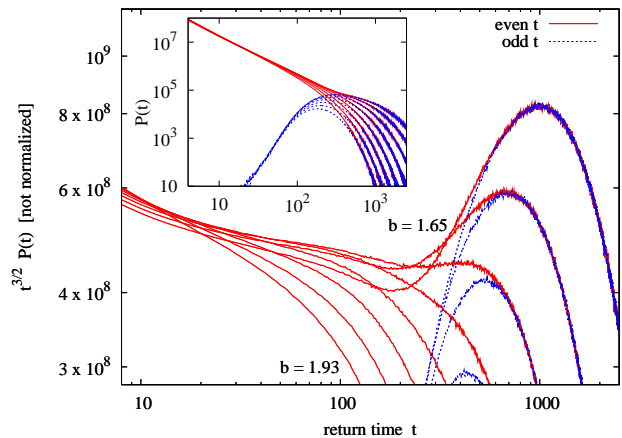, width=6.0cm, angle=270}
  \caption{(color online) Histograms of return times for walks on ER 
    graphs with $N=2\times 10^8$ and $L/N=1.5$. Shown is $t^{3/2}P(t)$ where
    $P(t)$ is the return probability. The inset is the plot of $P(t)$.
    The critical curve (third curve from above) 
    is the one which is flattest at $t\approx 100$. 
    Each curve corresponds to 
    a fixed bias $b$, with steps of 0.04. The upper (red) curves correspond 
    to even $t$, the lower (blue) ones to odd $t$.}
  \label{P_returntime}
  \end{center}
\end{figure}

More significant is the distribution of return times. In the inset of
Fig.~\ref{P_returntime} we plot (not normalized) histograms of return 
times $P(t)$ against $t$ for walks with various values of $b$, on 
the GC of an ER graph with $N=2\times 10^8$ and $L/N = 1.5$. 
For this graph $b_c = 1.73542$.  
The most striking observation is a very large 
even/odd effect for small $t$: Compared to the values for even $t$, $P(t)$
is nearly zero for odd $t$, an immediate consequence of the 
suppression of small loops. On a loopless tree, return times are always 
even. Conversely, we can assume that loops are negligible for even times 
for which $P(t\pm 1) \ll P(t)$. In this regime we should thus expect that 
the result for GW trees applies, {\it i.e.} $P(t) \sim t^{-3/2}$ at the critical 
bias. This prediction (which is the same as for unbiased 1-d 
%Brownian motion
random walks 
~\cite{peres::2006}) is in complete agreement with our data, as 
seen from Fig.~\ref{P_returntime}.
Fig.~\ref{P_returntime} also indicates the complicated finite-size behavior.
$P(t)$ is not convex at the 
critical point, and the usual finite size scaling ansatz (power law times 
homogeneous function) does not hold.

%\begin{figure}
%  \begin{center}
%  \epsfig{file=Fig-returntime-histo-2-3.ps, width=5.6cm, angle=270}
%  \caption{(color online) Replotting of the data shown in Fig.~\ref{P_returntime} 
%    as $t^{3/2}P(t)$ versus $t$. The critical curve (third curve from above) 
%    is the one which is flattest at $t\approx 100$.}
%  \label{P_returntime_detail}
%  \end{center}
%\end{figure}

\begin{figure}
  \begin{center}
  \epsfig{file=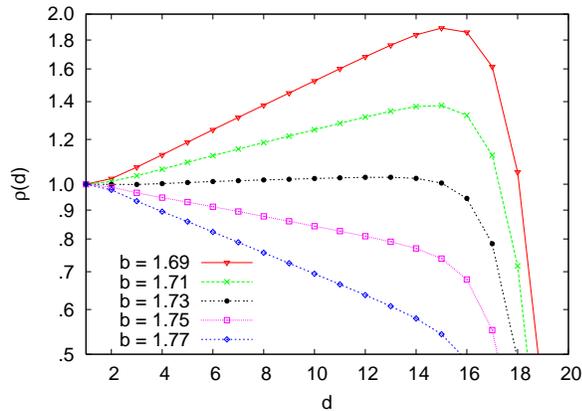, width=5.6cm, angle=270}
  \caption{(color online) Density of walkers against distance from the 
    target, for five values of $b$ close to criticality.}
  \label{dist-distrib}
  \end{center}
\end{figure}

Finally, if critical walks resemble unbiased 1-d Brownian motion, with 
reflecting boundaries at $d=0$ and at $d\approx \ln N$~\cite{peres::2006},
then the density $\rho(d)$ of walkers per launch, integrated over all times,
should be constant at $b=b_c$. 
Fig.~\ref{dist-distrib} shows that $\rho(d)$ is indeed flat,
to very high accuracy, for $b = 1.732\pm 0.002$. This agrees with the exact
$b_c$ within two standard deviations, and gives the most precise numerical 
verification of the theoretical prediction.

We have shown that walks on a random graph biased towards a given 
node show a localization/delocalization transition, corresponding to a 
transition from recurrent to transient behavior. If routing is noisy but
the noise level is below a critical threshold,
the walks are able to reach their destination in a finite amount of time.
This resilience of the walks can be taken into account while designing
routing strategies for the internet or other traffic problems with noisy
dynamics.  The localization/transition should 
also be related to a curious effect in quantum mechanics on 
random graphs with linearly rising potentials, where we expect a 
paradoxical ``unbinding" transition when the potential becomes too shallow.

Our numerical studies have only included ER graphs, mainly because these 
can be easily generated even with very large sizes. However, we expect 
the existence of a delocalization transition to be robust and to hold for 
any degree distribution with a finite second moment. For scale free graphs, 
$P(k) \sim 1/k^\gamma$ with $\gamma \leq 3$, the second moment diverges
with $N$, leading formally to $b_c=\infty$. For scale free graphs, the 
number of nodes within distance $d$ from the target does not grow exponentially, 
$\sim \mu^d$ as assumed in Eq.~(\ref{esceqn}). It rather grows super-exponentially 
as $\exp(\exp (d))$ \cite{chung,cohen::2003}. To compensate this, the critical
bias has to be distance-dependent, growing also super-exponentially with $d$. 
Alternatively, if we describe the finite-size behavior by an $N$-dependent 
but $d$-independent effective critical bias $b_c(N)$, then this has to 
grow as a power of $N$.

Finally, we should point out that our theoretical prediction for $b_c$ holds only for a 
very particular type of bias, given by Eq.~(\ref{jumps-ER}). For other biases we expect 
the transition to show the same scaling laws (as long as the bias strength is independent 
of $d$), although we can no longer predict the exact location of the phase transition. 
The same is true for graphs with many small loops, including those often observed in 
real world networks. If the bias strength increases (decreases) with $d$ beyond limit, 
then one has always (de)-localisation for graphs with a finite second-moment of the 
degree distribution and with exponential growth of the number of neighbors with distance.

We want to thank Maya Paczuski for numerous discussions 
and for  careful reading of the manuscript. VS would like to
thank Orion Penner for checking the calculations. PG thanks iCore
for financial support.

\bibliography{m8}

\end{document}